\documentclass[twocolumn]{openjournal}

\usepackage{graphicx}	
\usepackage{amsmath}	
\usepackage{amssymb}	
\usepackage{hyperref}
\usepackage{flexisym}	
\usepackage{gensymb}
\usepackage{natbib}
\usepackage{orcidlink}

\hypersetup{
     colorlinks   = true,
     linkcolor    = blue,
     citecolor    = blue
}

\begin{document}
\title{Analysis of optical spectroscopy and photometry of the type I X-ray bursting system UW CrB}

\author{\vspace{-1.5cm} M. R. Kennedy$^{1,2}$\orcidlink{0000-0001-6894-6044},
P. Callanan,$^{1}$,
P. M. Garnavich$^{3}$\orcidlink{0000-0003-4069-2817},
R. P. Breton$^{2}$\orcidlink{0000-0001-8522-4983},
A. J. Brown$^4$\orcidlink{0000-0002-3316-7240},
N. Castro Segura$^5$\orcidlink{0000-0002-5870-0443},
V. S. Dhillon$^{6,7}$\orcidlink{0000-0003-4236-9642},
M. J. Dyer$^6$\orcidlink{0000-0003-3665-5482},
S. Fijma$^{8}$,
J. Garbutt$^6$,
M. J. Green$^{9}$\orcidlink{0000-0002-0948-4801},
P. Hakala$^{10}$,
F. Jiminez-Ibarra$^{11}$\orcidlink{0000-0002-4634-1076},
P. Kerry$^6$,
S. Littlefair$^6$\orcidlink{0000-0001-7221-855X},
J. Munday$^{5}$\orcidlink{0000-0002-1872-5398},
P. A. Mason$^{12,13}$\orcidlink{0000-0002-5897-3038},
D. Mata-Sanchez$^{7,14}$\orcidlink{0000-0003-0245-9424},
T. Munoz-Darias$^{7,14}$,
S. Parsons$^6$\orcidlink{0000-0002-2695-2654},
I. Pelisoli$^5$\orcidlink{0000-0003-4615-6556},
D. Sahman$^6$\orcidlink{0000-0002-0403-1547}
}

\affiliation{$^{1}$Department of Physics, University College Cork, Cork, Ireland}
\affiliation{$^{2}$Jodrell Bank Centre for Astrophysics, School of Physics and Astronomy, The University of Manchester, M13 9PL, UK}
\affiliation{$^{3}$Department of Physics, University of Notre Dame, Notre Dame, IN, USA}
\affiliation{$^{4}$Departament de F\'{\i}sica, Universitat Polit\`{e}cnica de Catalunya, c/Esteve Terrades 5, 08860 Castelldefels, Spain}
\affiliation{$^{5}$Department of Physics, University of Warwick, Coventry CV4 7AL, UK}
\affiliation{$^{6}$Astrophysics Research Cluster, School of Mathematical and Physical Sciences, University of Sheffield, Sheffield, S3 7RH, UK}
\affiliation{$^{7}$Instituto de Astrof\'{i}sica de Canarias, E-38205 La Laguna, Tenerife, Spain}
\affiliation{$^{8}$Anton Pannekoek Institute for Astronomy, University of Amsterdam, Science Park 904, NL$-$1098 XH Amsterdam, the Netherlands}
\affiliation{$^{9}$Max-Planck-Institut f\"{u}r Astronomie, K\"{o}nigstuhl 17, D-69117 Heidelberg, Germany}
\affiliation{$^{10}$Finnish Centre for Astronomy with ESO (FINCA), Quantum, University of Turku, FI-20014, Finland}
\affiliation{$^{11}$School of Physics and Astronomy, Monash University, Clayton Campus, VIC 3800, Australia}
\affiliation{$^{12}$New Mexico State University, MSC 3DA, Las Cruces, NM 88003, USA}
\affiliation{$^{13}$Picture Rocks Observatory, 1025 S. Solano Dr. Suite D., Las Cruces, NM 88001, USA}
\affiliation{$^{14}$Departamento de Astrofísica, Univ. de La Laguna, E-38206 La Laguna, Tenerife, Spain}

\begin{abstract}
UW Coronae Borealis (UW CrB) is a low mass X-ray binary that shows both Type 1 X-ray and optical bursts, which typically last for 20 s. The system has a binary period of close to 2 hours and is thought to have a relatively high inclination due to the presence of an eclipse in the optical light curve. There is also evidence that an asymmetric disc is present in the system, which precesses every 5.5 days based on changes in the depth of the eclipse. In this paper, we present optical photometry and spectroscopy of UW CrB taken over 2 years. We update the orbital ephemeris using observed optical eclipses and refine the orbital period to 110.97680(1) min. A total of 17 new optical bursts are presented, with 10 of these bursts being resolved temporally. The average $e$-folding time of $19\pm3$s for the bursts is consistent with the previously found value. Optical bursts are observed during a previously identified gap in orbital phase centred on $\phi=0.967$, meaning the reprocessing site is not eclipsed as previously thought. Finally, we find that the apparent P-Cygni profiles present in some of the atomic lines in the optical spectra are due to transient absorption.
\end{abstract}

\section{Introduction}

UW Coronae Borealis (UW CrB), also known as MS 1603.6+2600, was discovered by the \textit{Einstein Observatory} Extended Medium Sensitivity Survey as an unusual binary with an orbital period of close to 2 hours \citep{1990ApJ...365..686M}. It was unusual in that it did not match the properties of any known cataclysmic variable (an interacting binary with a white dwarf primary; CV) and no Low Mass X-ray Binary (LMXB) had been discovered which had an orbital period in the 1-2 hour range. The optical spectrum of the source showed a strong blue continuum, Balmer, HeI, and HeII emission lines, and the Bowen blend emission feature at 4640 \AA, and optical light curve showed variations on a timescale of 2 hours (hence the orbital period estimate). However, the general shape and orbital phase of these variations were found to change from night to night. \cite{1998A&A...333..540H} confirmed the variations in the optical light curve, and also ruled out a non-magnetic CV classification for UW CrB. They preferred an LMXB classification for UW CrB, despite its short orbital period. \cite{2001ApJ...561..938M} discovered a candidate type I X-ray burst from UW CrB using the \textit{Advanced Satellite for Cosmology and Astrophysics} (\textit{ASCA}), lending substantial weight to the LMXB classification with a neutron star (NS) primary.

Type I X-ray bursts are indicative of a NS primary as they are thought to arise during brief periods of thermonuclear runaway on the surface of the NS. Type I X-ray bursts are typically short in duration, with a sharp rise lasting between $1-10$ s and slow decay lasting between 10 s and a few minutes (see \citealt{2013PrPNP..69..225P} for a review). The observed bursts in UW CrB match this description perfectly, with rise times of $<5$ s and a decay time of $\sim20$ s.

However, given the candidate nature of the burst presented by \cite{2001ApJ...561..938M}, final classification of the system had to wait until the studies of \cite{2004ApJ...608L.101H} and \cite{2004RMxAC..20..211M}. Both of these works discovered that UW CrB also produces optical bursts that last between 15$-$30 s and increase the optical flux from the system by a factor of 2$-$3. They proposed that these optical bursts are counterparts to the X-ray bursts, and are produced by reprocessing of the X-ray bursts by material in the system. These type I optical bursts are not unique to UW CrB, and have been seen in a handful of other systems (MXB 1735-44, \citealt{1978Natur.274..567G}; V801 Arae and UY Volantis, \citealt{1990A&A...227..105S}; Aquila X-1, \citealt{1997ApJ...491L..89R}; GS 1826-24, \citealt{1998MNRAS.298..497H}; and others). Determination of the exact correlation between the optical and X-ray bursts seen in UW CrB has been hampered by the lack of any simultaneous X-ray and optical observations.

Since 2004, multiple optical observations of UW CrB have revealed more about the nature of the binary. \cite{2008ApJ...685..428M} refined the orbital period to 110.97671 min, and discovered that the nightly variations in the shape and width of the eclipse varied periodically with a 5.5 day period. They proposed that such behaviour can be explained by an asymmetric accretion disc in the system which precesses around the NS once every 5.5 days. \cite{2009MNRAS.394..892H} confirmed the super-orbital period of 5.5 days, and also analysed 11 more optical bursts, finding a similar duration to \cite{2004ApJ...608L.101H}. \cite{2012AJ....144..108M} presented analysis of an additional 9 optical bursts, and combined the optical bursts from their work with the bursts presented in \cite{2004ApJ...608L.101H} and \cite{2009MNRAS.394..892H}. They found that no optical bursts had been observed during an orbital phase gap centred on orbital phase 0.967 and with a width of $\delta \phi =0.210$, and suggested that this was due to an eclipse of the reprocessing site by the companion star. Here, and in the rest of this paper, an orbital phase of zero occurs when the companion star is at inferior conjunction. Finally, and most recently, \citet{2023MNRAS.526L.149F} presented ultraviolet spectra of UW CrB taken with the Cosmic Origins Spectrograph onboard the Hubble Space Telescope in 2011. In these data, a transient P Cygni profile, indicative of an outflow, is visible. These authors consider in detail whether a disc wind would be expected in a system with parameters consistent with UW CrB, while also mentioning that such line profiles could also arise from irradiation driven evaporation of the companion star, or an outflow coming from the interaction point (the ``hot spot'') between the ballistic stream and the accretion disc.

Here, we present new optical photometry taken with the intent to extend the optical ephemeris from \cite{2012AJ....144..108M}, and also to observe and characterise more optical bursts. We also present optical spectroscopy of UW CrB, the first taken in 20 years.

\section{Observations}
Below, we detail observations taken of UW CrB between 2014 and 2019.

\subsection{VATT \& LBT Photometry}
\begin{figure*}
	\centering
	\includegraphics[width=\textwidth]{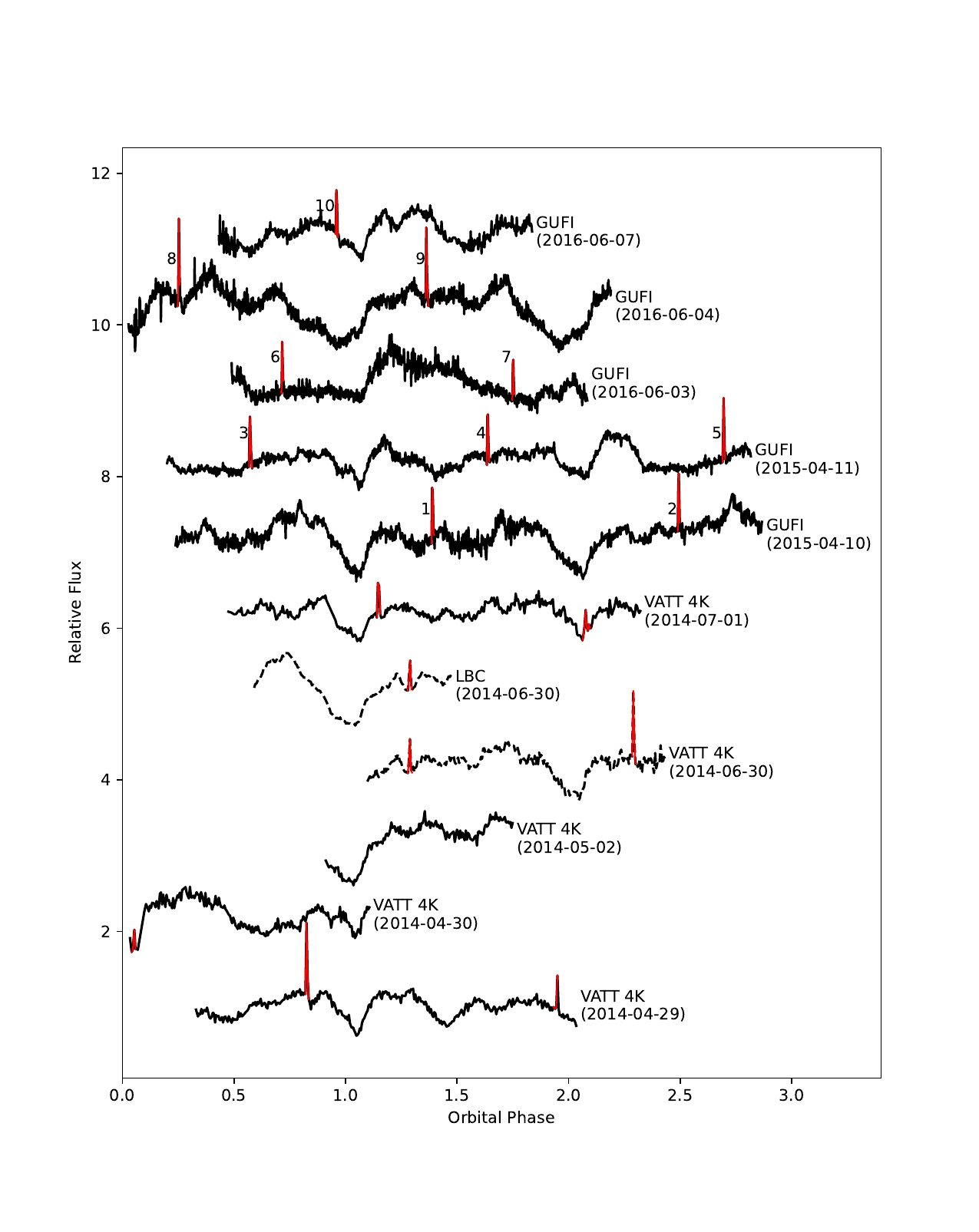}
	\caption[Light curve of UW CrB]{The individual light curves for each observation listed in Table~\ref{tab:uwcrb_observations} (with the exception of the HiPERCAM data). The light curves have been offset from each other by a constant factor, and are plotted in ascending order, with the bottom light curve taken on 2014-04-29 and the top light curve taken on 2016-06-07. The data plotted in red are optical bursts whose identification is described in the text. The 10 bursts taken using the GUFI instrument, which are numbered 1-10, are analysed in depth in Section~\ref{sec:Bursts}. Note that the data taken on 2014-06-30 from the VATT and LBT (shown as dashed black lines) overlap, and show the same bursts.}
	\label{fig:uwcrb_lc}
\end{figure*}

Optical photometry of UW CrB were carried out over 10 nights between 2014-2016. Table~\ref{tab:uwcrb_observations} shows the full details of individual observations. The Vatican Advanced Technology Telescope (VATT) 2014 observations were taken using the VATT4K CCD with a 20 s exposure time and additional 14 s overhead, and with a SDSS-g\textprime\ filter inserted. The Large Binocular Telescope (LBT) 2014 observations were taken using the Large Binocular Camera (LBC, \citealt{LBC1}) mounted on the DX (right side) of the telescope with a Johnson V filter inserted. The VATT 2015 and 2016 observations were taken using the Galway Ultrafast Imager (GUFI, \citealt{GUFI2}) with varying exposure times, without any filter so as increase the S/N of the detection of the source in each frame. The advantage of using GUFI over the VATT4K is that GUFI is a frame transfer CCD and has negligible read out time. The data were bias corrected and flat fielded using \textsc{iraf} \footnote{IRAF is distributed by the National Optical Astronomy Observatories, which are operated by the Association of Universities for Research in Astronomy, Inc., under cooperative agreement with the National Science Foundation.}, and simple aperture photometry with one comparison star was used to extract the light curve. Figure~\ref{fig:uwcrb_lc} shows the light curve from each individual run.

\begin{table}
	\centering
	\caption{Details of the various optical observations of UW CrB between 2014 and 2019.}
	\begin{tabular}{l c c c c}
		\hline
        Date (UT)                          & Instrument                & Cadence   & No.                 &Phase\\
		&                          & (s)       & of Frames           &Coverage  \\
		\hline\hline
		2014-04-29	& VATT4K                    & 34        & 335                 & 0.3-2.3\\
		2014-04-30	& VATT4K                    & 34        & 212                 & 0.3-1.1\\
		2014-05-02	& VATT4K                    & 34        & 166                 & 0.9-1.7\\
		2014-06-30	& VATT4K                    & 34        & 270                 & 0.1-1.4\\
		2014-06-30	& LBC                       & 34        & 169                 & 0.6-1.5\\
		2014-07-01	& VATT4K                    & 34        & 366                 & 0.5-2.3\\
		2015-04-10	& GUFI                      & 15        & 1130                & 0.2-2.9\\
		2015-04-11	& GUFI                      & 15        & 1151                & 0.2-2.8\\
		2016-06-03	& GUFI                      & 10        & 1020                & 0.5-2.1\\
		2016-06-04	& GUFI                      & 10        & 1387                & 0.0-2.2\\
		2016-06-07	& GUFI                      & 17        & 583                 & 0.4-1.8\\
		2019-07-08  & HiPERCAM                  & 0.5       & 30101     & 0.0-1.75\\
		\hline
	\end{tabular}
	\label{tab:uwcrb_observations}
\end{table}

\subsection{LBT Spectroscopy}
While LBC on the DX mirror of the LBT was taking photometry of UW CrB, the Multi-Object Dual Spectrographs (MODS1; \citealt{PoggeMODS}) on the SX mirror (left side) was taking simultaneous spectroscopy of UW CrB. MODS1 in the dual grating mode covers a wavelength range from 320 nm to 1 $\mu$m, divided into red and blue channels separated by a dichroic at 560 nm. The exposure for each spectrum was 60 s and the time between the start of consecutive exposures averaged 120 s. The readout time of the red channel on the MODS instrument is faster than on the blue channel, meaning the red set of spectra finished before the blue set, leading to a slightly different time coverage in the separate channels. Fifty spectra were obtained from each channel between 06:18 UT and 7:54 UT. The data reduction tasks were carried out using \textsc{iraf}.

\subsection{HiPERCAM Photometry}
\begin{figure*}
	\centering
	\includegraphics[width=\textwidth]{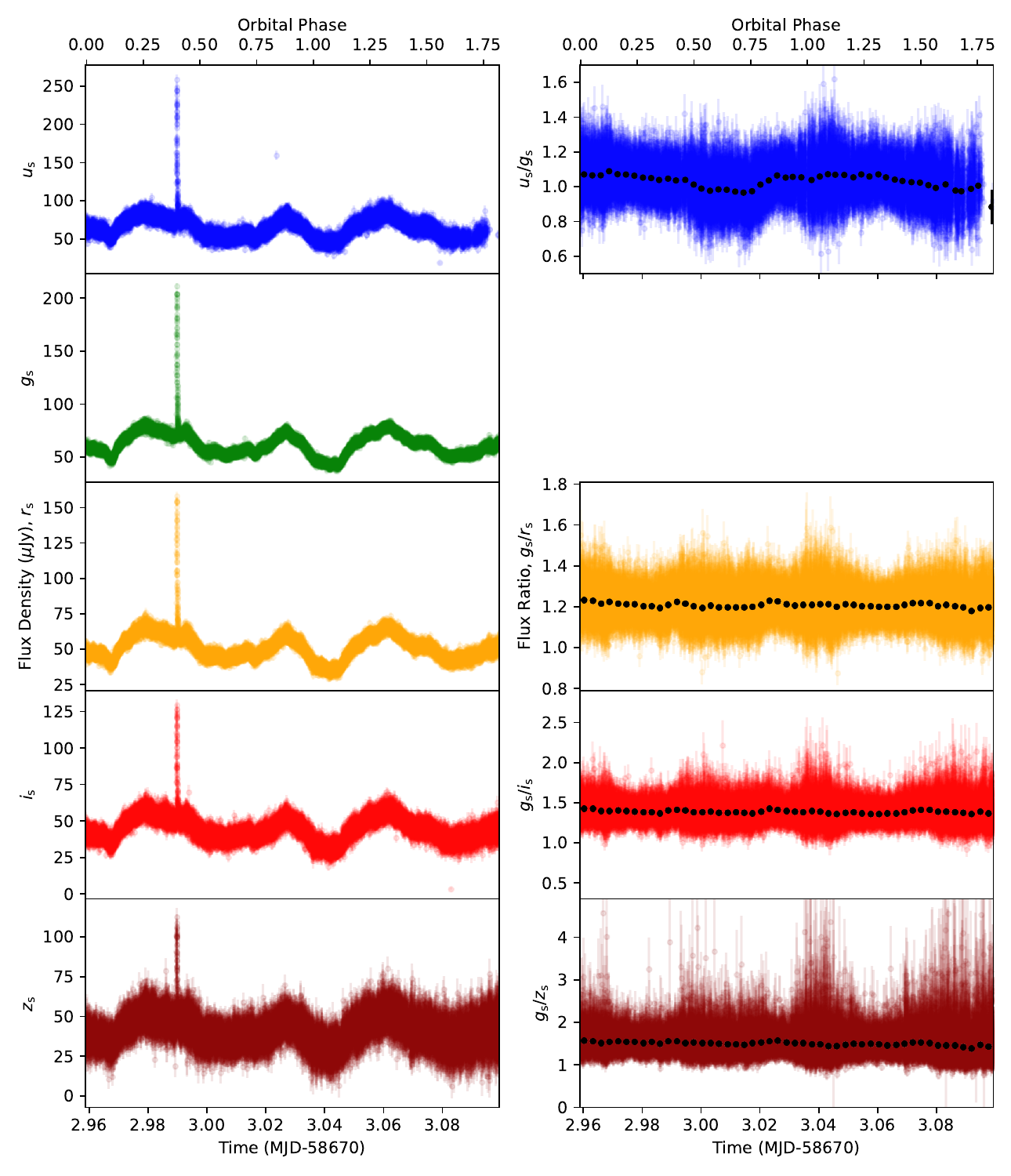}
	\caption{HiPERCAM light curve of UWCRB. \textit{Left column:} Each panel represents a different filter, with $u_{\rm s}$ on top and $z_{\rm s}$ on bottom. There is a single burst detected in all filters. The time resolution of this light curve is 0.5 s, a substantial increase on the data shown in Figure~\ref{fig:uwcrb_lc}. The orbital phase of the binary is given as the top axis, and was calculated using the ephemeris presented in this paper. \textit{Right column:} The ratio of fluxes in each band. The top panel shows the $u_{\rm s}$ flux relative to the $g_{\rm s}$ flux, while the other panels show the $g_{\rm s}$ flux relative to the corresponding band in the left column. This is included to show the small variation in the colour of the system over an orbital period. The black data points in each panel are the median combined points after splitting the data into 50 equally spaced time bins.}
	\label{fig:uwcrb_hcam}
\end{figure*}

UW CrB was observed using the quintuple beam photometer HiPERCAM \citep{HCAM} mounted on the 10.4m Gran Telescopio Canarias, starting at 2019-07-08 23:00:37 and lasting for 2 hours and 20 minutes. HiPERCAM provides five simultaneous bands of optical photometry with a dead time $\sim$ 7.8ms. The data were acquired with Super-SDSS $u_{\rm s}$, $g_{\rm s}$, $r_{\rm s}$, $i_{\rm s}$ and $z_{\rm s}$ filters, which are filters that cover the same wavelength range as the traditional SDSS \textit{u\textprime}, \textit{g\textprime}, \textit{r\textprime}, \textit{i\textprime}, and \textit{z\textprime}\ filters \citep{2010AJ....139.1628D}, but with a higher throughput \citep{HCAM}. The exposure time for each frame was 0.5 s, with 4x4 on-chip binning used for each CCD. The data were reduced using the purpose-built HiPERCAM pipeline (as described in \citealt{HCAM}) using biases and flat field observations from that same night, and flux calibrated using the SDSS magnitudes of a nearby star located at $(\alpha=241.442353^{\rm o},\delta=+25.869876^{\rm o})$. The light curve of these observations are presented on their own in Figure~\ref{fig:uwcrb_hcam}, along with the ratio of fluxes of each band relative to the $g_{\rm s}$. This is provided to demonstrate that the colour of the system does not vary significantly over the orbital period.

\subsection{Hubble Space Telescope Cosmic Origins Spectrograph data}
We additionally include thee epochs of data taken of UW CrB using the Cosmic Origins Spectrograph (COS, \citealt{HSTCOS}) on the Hubble Space Telescope (HST). Epoch one covered 02:15 to 09:12 UT on September 1, 2011, and was first presented in \citet{2023MNRAS.526L.149F}. Epoch two covered several windows between August 21 and August 22 2023. Epoch three covered 18:04 UT to 20:18 UT on February 21, 2024. Detailed analysis of the second and third epochs are on going, but here we include the timing of the bursts observed in these data when discussing the phases at which optical bursts from UW CrB are observed. While each epoch was taken using HST/COS, the 2011 data, in which an optical burst is present, were obtained using the G160M grating (spanning $\sim$1360-1775 \AA), and the 2023 and 2024 data were obtained using the G140L grating (spanning $\sim$1100-2290 \AA). Light curves were generated using data taken within a wavelength range of 1415-1770 \AA\ for the former and 1170-2000 \AA\ for the latter. Here the geocoronal Lyman-$\alpha$ (1208-1225 \AA) and O I (1298 to 1312 \AA) line profiles were masked, as well as low signal-to-noise regions near the edges of the gratings

\section{Results}
\subsection{Orbital Ephemeris}
\begin{figure}
	\centering
	\includegraphics[width=\columnwidth]{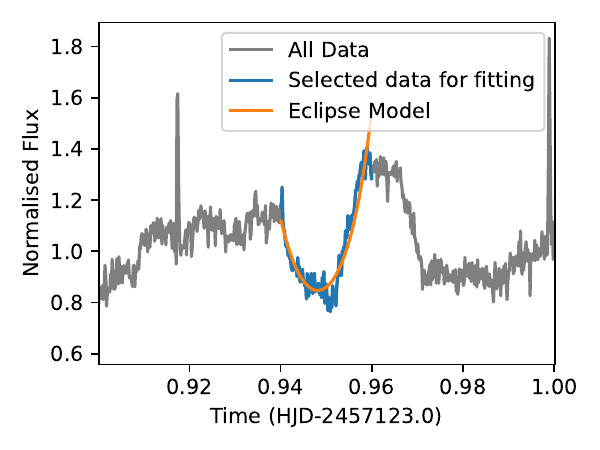}
	\caption{An example of the quadratic fit to eclipse data. The grey line shows the normalised flux, while the highlighted blue data are those that were selected as the eclipse data. The orange line shows the quadratic fit to these data.}
	\label{fig:example_eclipse}
\end{figure}

The data presented fully sampled 13 eclipses and partially sampled 3 eclipses. Table~\ref{tab:uwcrb_midecl_times} lists the mid times of these eclipses. Since the eclipse changes profile not only from night to night, but also from orbit to orbit, it is impossible to pick a single feature in the eclipse that could serve as a reference point for calculating the mid-eclipse time. As such, we identified the eclipse ingress and egress in each light curve, fit a quadratic function to the data spanning this range, and set the eclipse mid-time to the centre of this quadratic fit. An example of this is shown in Figure~\ref{fig:example_eclipse}.

The corresponding eclipse numbers were calculated using the orbital ephemeris presented in \cite{2012AJ....144..108M}. These eclipse times were added to the 32 mid-eclipse times from \cite{2012AJ....144..108M} and the 24 mid-eclipse times from \cite{2008ApJ...685..428M}. All 72 mid-eclipse times were fit to determine an accurate ephemeris. The resulting linear ephemeris was

\begin{equation} 
    T_{mid}(HJD) = 2453118.8390(4) + 0.077067223(9) E \label{uwcrb_lin_eph}.
\end{equation}

\begin{figure}
	\centering
	\includegraphics[width=\columnwidth]{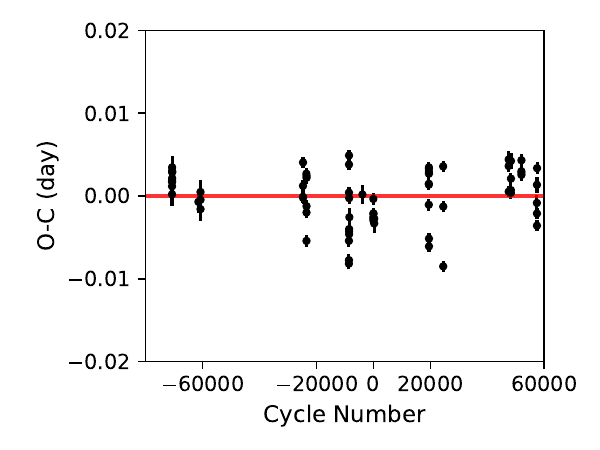}
	\caption[O$-$C for UW CrB]{The observed-calculated times of mid-eclipse for UW CrB. The cycle number was calculated using Equation \ref{uwcrb_lin_eph}.}
	\label{fig:uwcrb_oc}
\end{figure}

The O-C curve for this linear ephemeris is shown in Figure~\ref{fig:uwcrb_oc}, and this ephemeris was used for calculating the orbital phases shown in Figure~\ref{fig:uwcrb_lc}. The $\chi^{2}$ value for this fit was 1388 for 70 degrees of freedom. A quadratic fit to the mid-eclipse times was also tested, which gave a quadratic ephemeris of 

\begin{equation}\label{uwcrb_quad_eph}
\begin{split}
T_{mid}(HJD) = {} & 2453118.8377(4) + 0.077067235(9) E\\
          		 & + (8(2) \times 10^{-13}) E^{2}.
\end{split}
\end{equation}

The $\chi^{2}$ value for this fit was 1213 for 69 degrees of freedom. While the quadratic ephemeris provides an improvement in $\chi^2$, there is still significant scatter in the O-C constructed using this ephemeris. This is most likely related to the known variation in the mid-eclipse time across the 5.5 day period which has been attributed to a precessing accretion disc in the system. Without knowledge of the disc phase during our observations (none of our observations cover a continuous 5.5 days), it is impossible to tell if the additional term in the quadratic ephemeris is due to a change in the orbital period, or just related to minimising this scatter. Performing an F-test using the variances of the residuals of both ephemerides gives a P value of 0.22, which suggests we cannot reject with high confidence the null-hypothesis that the linear model is a good fit to the data, and given that the orbital period in both ephemerides agree within 1$\sigma$, we adopt the linear ephemeris for the rest of this paper. 

\begin{table}
	\centering
	\caption{Mid-eclipse times for the 16 eclipses in the new data.}
	\begin{tabular}{l l l}
		\hline
        Eclipse No.     & Mid-eclipse Time  & Error\\
		            	& (HJD)             & (day)\\
		\hline\hline
		47465		    & 2456776.8384	    & 0.0007\\
        47477		    & 2456777.764	    & 0.001\\
        47505		    & 2456779.9180	    & 0.0007\\
        48269		    & 2456838.7972	    & 0.0007\\
        48270		    & 2456838.8746	    & 0.0007\\
        48282		    & 2456839.8008	    & 0.0007\\
        48283		    & 2456839.880	    & 0.001\\
        51955		    & 2457122.8696	    & 0.0007\\
        51956		    & 2457122.9465	    & 0.0007\\
        51968		    & 2457123.8728	    & 0.0007\\
        51969		    & 2457123.9481	    & 0.0007\\
        57403		    & 2457542.728	    & 0.001\\
        57415		    & 2457543.655	    & 0.001\\
        57416		    & 2457543.7286	    & 0.0007\\
        57417		    & 2457543.8042	    & 0.0007\\
        57455		    & 2457546.7397	    & 0.0007\\
		\hline
	\end{tabular}
	\label{tab:uwcrb_midecl_times}
\end{table}

\subsection{Optical Spectrum}

\begin{figure*}
	\centering
	\includegraphics[width=\textwidth]{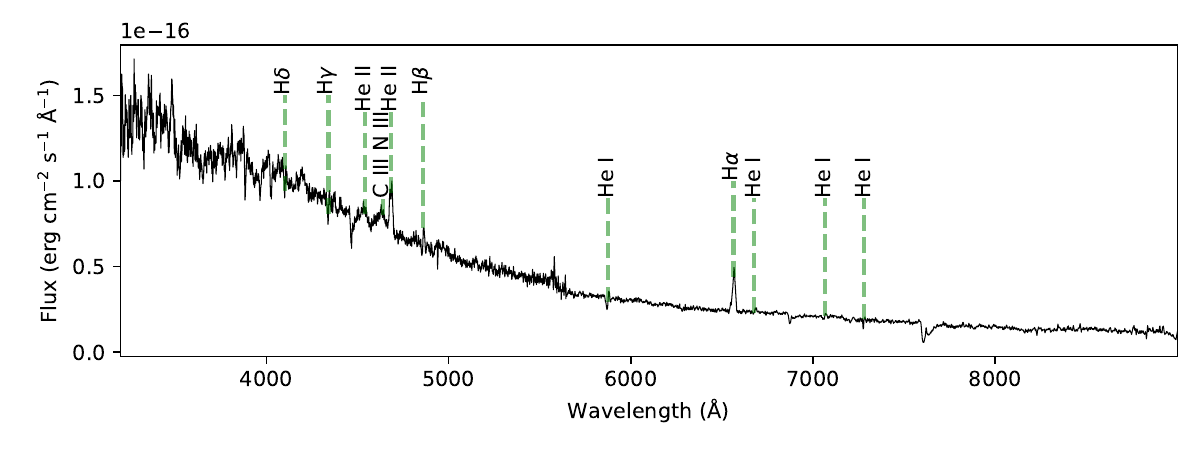}
	\caption[Average optical spectrum of UW CrB]{The average spectrum of UW CrB taken using the MODS instrument. The most prominent emission lines are marked. The change in the signal to noise ratio at $\sim$ 5700 \AA\ is where the blue detector ends and the red detector begins.}
	\label{fig:uwcrb_spec}
\end{figure*}

\begin{figure}
	\centering
	\includegraphics[width=\columnwidth]{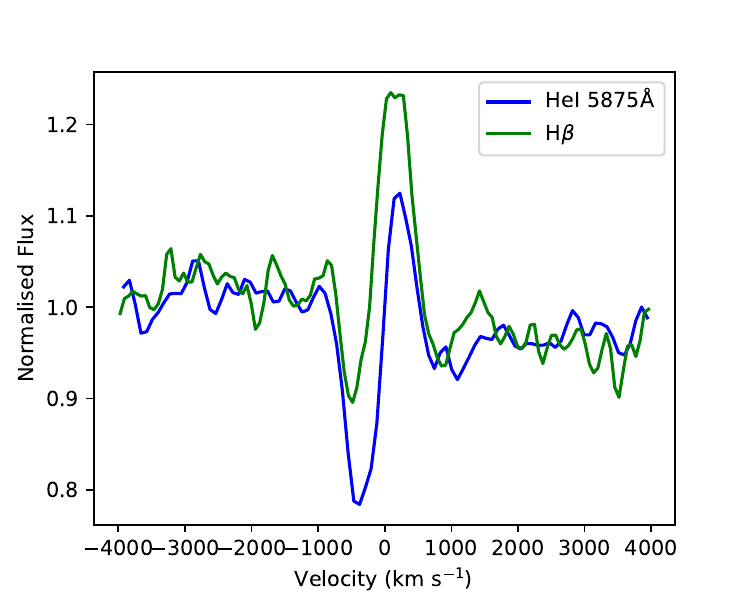}
	\caption[P-Cygni lines in UW CrB]{The average line profiles of H$\beta$ (green) and HeI 5875 \AA\ (blue), showing strong blue shifted absorption and red shifted emission.}
	\label{fig:uwcrb_pcyg}
\end{figure}

The average optical spectrum is shown in Figure~\ref{fig:uwcrb_spec}. The most prominent emission and absorption lines are marked, which correspond to H$\alpha$, H$\beta$, a series of HeI and HeII lines, and the Bowen blend structure (CIII and NIII) at 4640 \AA. The most interesting features of the optical spectrum are the absorption features seen in the HeI lines and in H$\beta$. Figure~\ref{fig:uwcrb_pcyg} shows the time averaged line profile of HeI 5875 \AA\ alongside H$\beta$. The average line profiles resemble a P Cygni line profile, which are characterised by blue-shifted absorption components and a red-shifted emission component. Such line profiles, first proposed by \citet{1929MNRAS..90..202B}, are indicative of an outflow of material from a system and have been studied in detail in the context of LMXBs and CVs. They often vary in strength (e.g., depending on the flaring activity during the outburst), can be transient in nature (vanishing over shorter than hourly timescales), and can be difficult to interpret (see for example \citealt{2016Natur.534...75M} and \citealt{2019ApJ...879L...4M}).

\begin{figure}
	\centering
	\includegraphics[width=\columnwidth]{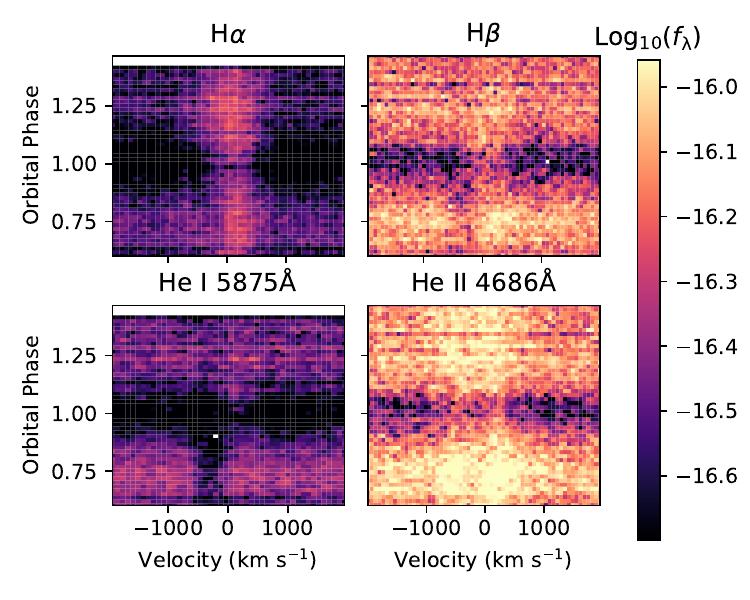}
	\caption[The trailed optical spectrum of UW CrB]{The trailed spectra of H$\alpha$, H$\beta$, HeII and HeI (going clockwise from top left). These trailed spectra show how the intensity (represented by the colour) of the lines varies over the orbital phase. In particular, all the lines experience a drop in intensity during eclipse. HeII shows a strong double peaked nature throughout the binary orbit, while HeI shows transient absorption.}
	\label{fig:uwcrb_trail}
\end{figure}

To understand the line profile better, trailed spectra for H$\alpha$, H$\beta$, HeI 5875 \AA, and HeII 4686 \AA\:are given in Figure~\ref{fig:uwcrb_trail}. These plots highlight the transient nature of the absorption feature in several of these lines. From the H$\alpha$ plot, we can see that the emission peak is largely confined to positive velocities. HeII, on the other hand, shows clear blue and red emission components both before and after eclipse. HeI shows absorption at negative velocities prior to the eclipse, and very little absorption or emission after the eclipse.

\begin{figure}
	\centering
	\includegraphics[width=\columnwidth]{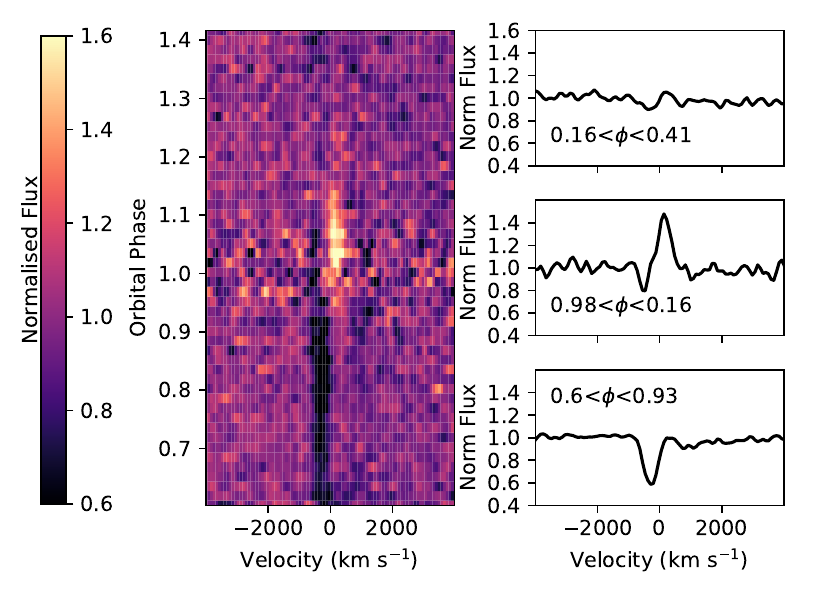}
	\caption[The trailed HeI line in UW CrB]{The left panel shows the trailed spectrum around HeI 5875\AA. The right panels show the average spectrum for selected orbital phases, highlighting when the line displayed strong absorption, strong emission and just continuum. The spectra here have been smoothed by convolving the spectrum with a gaussian with a $\sigma=0.9$\AA.}
	\label{fig:uwcrb_trail_heI}
\end{figure}

Figure~\ref{fig:uwcrb_trail_heI} shows the normalised trailed spectrum around HeI 5875\AA. When compared to Figure 2 of \citealt{2016Natur.534...75M}, which shows the standard behaviour for P Cygni profiles in LMXBs where both emission and absorption components are visible simultaneously, it becomes apparent that the behaviour of the line profile of HeI in the spectrum of UW CrB is not that of a conventional P Cygni line, where the absorption component is visible without any emission counterpart for orbital phase $0.6-0.93$. Then, around the time of optical eclipse, the line undergoes a reversal, with the absorption feature becoming weaker, and a strong, red-shifted emission component develops in the line profile. Finally, at orbital phase 0.16, both features disappear (or, at the very least, become significantly weaker, as shown by the top average spectrum shown in Figure~\ref{fig:uwcrb_trail_heI}). The strong emission seen around orbital phase 0 is probably due to a weakening of the continuum as the accretion disc is eclipsed, allowing for the line to become detectable, as opposed to an actual strengthening of the emission line itself. This suggests that the emission feature is present for the whole orbit and, outside of eclipse, it is drowned out by the continuum. Thus, we only treat the absorption feature as transient.

\subsubsection{Optical Burst Spectrum}
\begin{figure}
	\centering
	\includegraphics[width=\columnwidth]{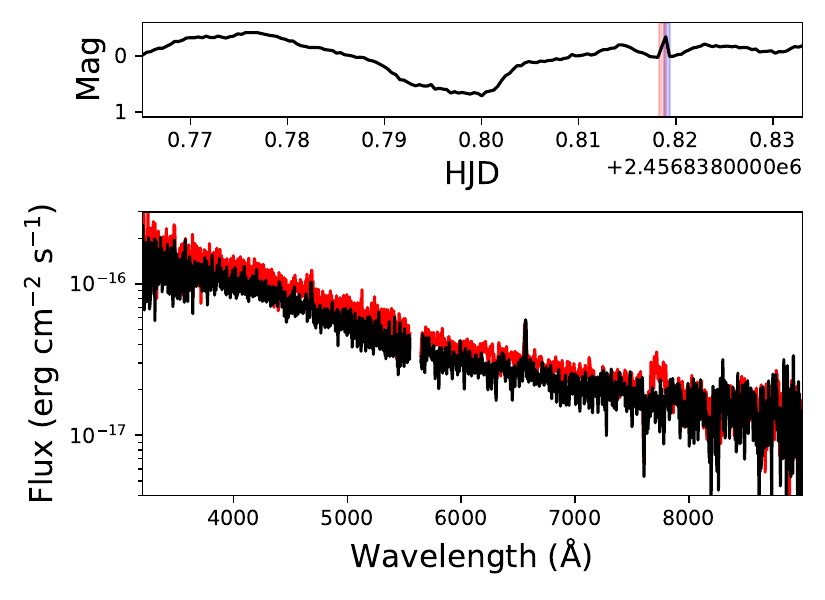}
	\caption[Spectrum of an optical burst in UW CrB]{The top panel shows the LBC light curve, focused on the time around the optical burst. The transparent red region highlights the time when the red spectrograph of MODS1 was exposed, while the transparent blue region highlights the time when the blue spectrograph was exposed. The main panel shows the entire optical spectrum observed during the burst (red) versus the average of the spectra taken directly before and after the burst (black).}
	\label{fig:burst_spec}
\end{figure}

The top panel of Figure~\ref{fig:burst_spec} shows the LBC light curve around the time of the optical burst seen in the LBT data. Fortunately, both the blue and red sides of the spectrograph were exposed when this burst occurred. This observed spectrum is shown in the main panel of Figure~\ref{fig:burst_spec}, and has a stronger continuum than the spectra taken before and after the burst. However, no changes in any of the emission lines is detectable, with H$\alpha$ being the best resolved emission feature. The residual spectrum (burst spectrum - average spectrum) is consistent with having a spectral index of 0 (that is, the spectrum is flat). This means we cannot get any temperature constraints on the burst from the optical spectrum. The detection of the optical burst spectrum is promising, and suggests a larger telescope will be able to obtain higher signal-to-noise burst spectra which will help constrain the site of the optical reprocessing. The physical origin of the feature around 7700\AA\: in the optical burst spectrum is unknown, but is visible in several spectra throughout the orbit, suggesting it is not related to the burst.

\subsection{Optical Bursts}\label{sec:Bursts}
A total of 18 optical bursts were detected in all of the observations taken of UW CrB - 17 with using the VATT, 1 using the LBT, and 1 using the GTC. For the VATT and LBT data, these bursts were identified in the observed light curves by looking for points which were more than 3$\sigma$ away from the median of the surrounding 40 points, where $\sigma$ is the standard deviation of the 40 points. These bursts were then inspected by eye as a final confirmation step. Seven of these bursts were detected using the VATT4K instrument with a temporal resolution of 35s, one of which was also detected in the LBC observations. Since the optical bursts have been observed to have an $e-$folding time of $~20$s, this means the VATT4K observations were unable to resolve bursts, with each burst only recorded as a single data point in the light curves. Even though the structure of these bursts cannot be explored, the orbital phases at which they were detected is important, and will be discussed in Section \ref{sec:PhaseOfBursts}.

\subsubsection[Modelling Optical bursts]{Modelling the optical bursts I - GUFI}
There were ten bursts detected using the GUFI instrument with a temporal resolution of 15s or better. These bursts were fit using an instantaneous rise and exponential decay model, as done previously by \cite{2004ApJ...608L.101H} and \cite{2009MNRAS.394..892H}, after subtraction of a linear fit to the data around the burst to account for the orbital variation in the light curve. This model has the form

\begin{equation}
    F(t) = A\: e^{\left( \frac{t-t_0}{\tau} \right)} \label{eqn:BurstModel},
\end{equation}

where $A$ is the peak amplitude of the burst, $t_0$ is the start time of the burst, and $\tau$ is the $e-$folding time of the bursts. The fitting was performed using \textsc{Multinest} (\citealt{2008MNRAS.384..449F}; \citealt{2009MNRAS.398.1601F}; \citealt{2013arXiv1306.2144F}) implemented in \textsc{Python} using \textsc{Pymultinest} \citep{2014A&A...564A.125B}. $A$ and $t_{0}$ were both sampled using flat priors, while a Gaussian prior centered on 20 s with $\sigma=10$ s was assumed for $\tau$. We additionally fit the 7 bursts originally presented in \citet{2015AcPPP...2...50M}, since these bursts were never analysed in this way.

\begin{figure}
	\centering
	\includegraphics[width=1.0\columnwidth]{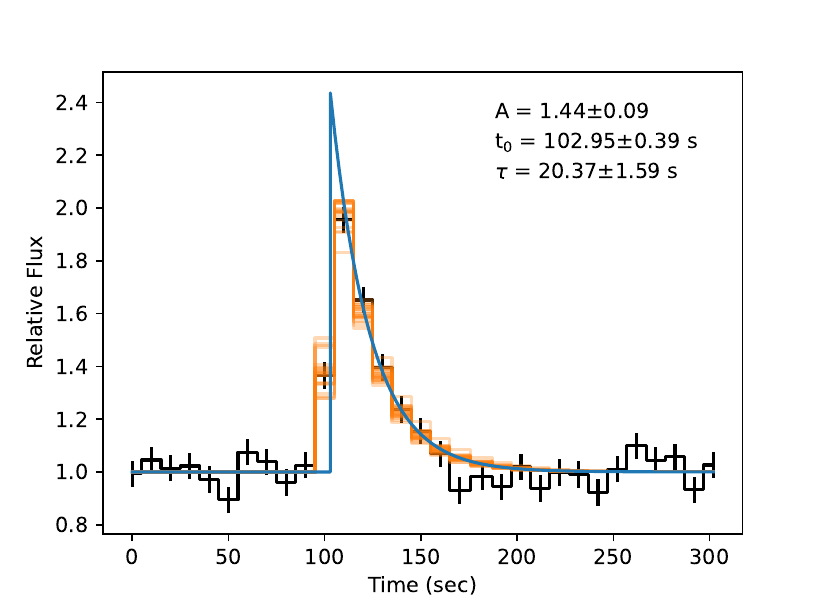}
	\caption{The results of fitting Equation~\ref{eqn:BurstModel} to burst 9. The best-fit parameters and errors are given in the top right, and the blue line shows the best-fit model when sampled with a time resolution of 0.01 s. The orange values show 100 models binned to the same temporal resolution as the data (black with associated errors) with parameter values drawn from the posterior distributions.}
	\label{fig:burst_nine}
\end{figure}

Figure~\ref{fig:burst_nine} shows the results of this fitting when applied to the burst numbered 9 in Figure~\ref{fig:uwcrb_lc}, with the time of this plot being an arbitrary cut in the data around the burst. The start time of the burst is measured to an accuracy 0.38 s based on reproducing the correct flux levels in the first and second bins of the burst, while the $e-$folding time is constrained within 2 s (all errors are quoted at the $1\sigma$ level). 

\begin{figure*}
	\centering
	\includegraphics[width=1.0\textwidth]{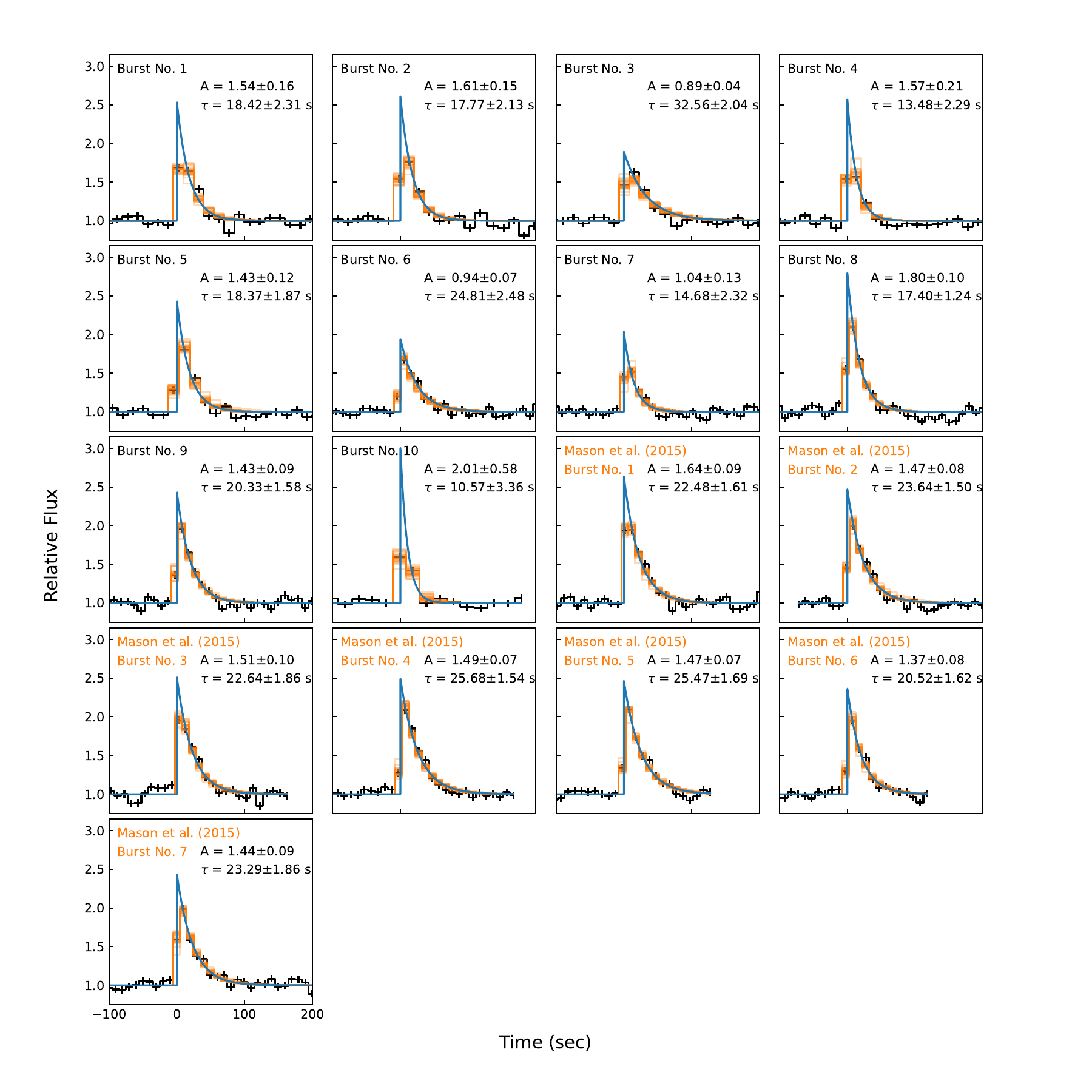}
	\caption[Model bursts]{The 10 optical bursts observed with the GUFI instrument and 7 from \citet{2015AcPPP...2...50M} (in black) along with the resulting model from fitting Equation \ref{eqn:BurstModel} (blue). The orange lines here represent 100 models drawn from the posterior distributions of the parameters of each burst so as to give an idea of the error in the model.The value of $t=0$ corresponds to the peak amplitude in the model burst. The error in the amplitude of Burst 10 is significantly higher than the other bursts due to the longer cadence of these data.}
	\label{fig:model_bursts}
\end{figure*}

The results of applying this method to all of the 10 bursts observed using GUFI and the 7 bursts from \citet{2015AcPPP...2...50M} can be seen in Figure~\ref{fig:model_bursts}. In this plot, the best-fit $t_0$ has been subtracted from data such that all of the plots have $t=0$ located at the peak of the burst. There is a large scatter in the measured $e-$folding times, ranging from $11\pm3$ s at the shortest up to $32\pm3$ s. The observed $e-$folding times are in good agreement with the range of values between 15$-$28 s and 15$-$27 s reported in \cite{2004ApJ...608L.101H} and \cite{2009MNRAS.394..892H}.

\subsubsection{Modelling the optical bursts II - HiPERCAM}
\begin{figure}
	\centering
	\includegraphics[width=1.0\columnwidth]{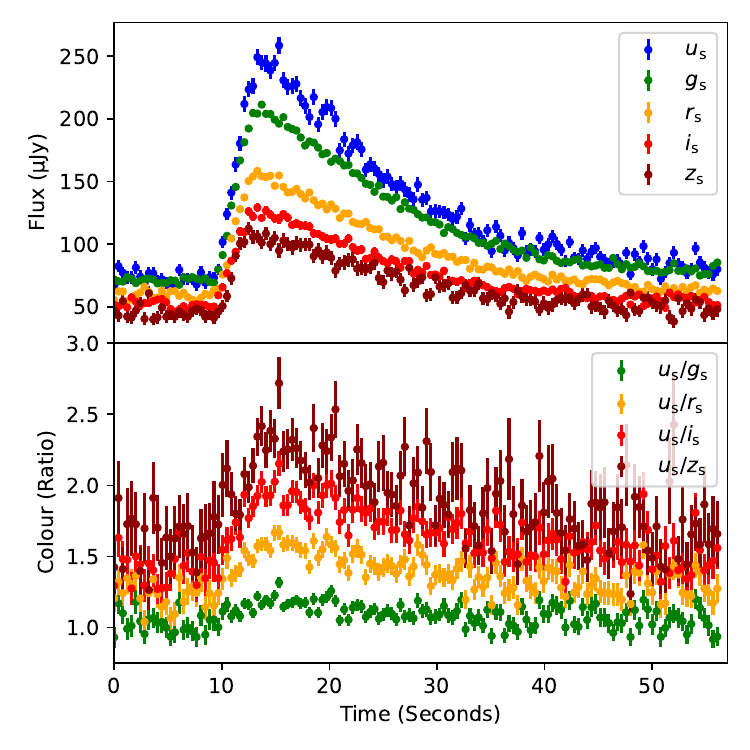}
	\caption[HCAM Burst]{The optical burst observed using HiPERCAM in the 5 SDSS bands, plotted in calibrated flux density (top), and as a ratio of the $u_{\rm s}$-band to each separate band (bottom). Here, the absolute time of the first datum has been subtracted off to give clarity to the duration of the burst rise and decay time.}
	\label{fig:HCAM_Burst}
\end{figure}

The burst detected using HiPERCAM is shown in Figure~\ref{fig:HCAM_Burst}, and represents a substantial increase in data quality over the rest of the data present in this paper. As such, we attempted to fit this burst with a physically motivated model, as opposed to the simple exponential of Equation~\ref{eqn:BurstModel}.

Creating a self-consistent model where an X-ray burst occurs on the surface of the neutron star, then heats the elliptical disc and the exposed surface of the companion star is not feasible with current binary star modelling programs. There have been previous attempts to model the relation between the energy and duration of an X-ray burst and its associated optical counterpart. An excellent example of this is given in \citet{2006ApJ...648.1156H}, who assume a Gaussian transfer function to estimate the change in temperature and the area of the reprocessing site. Here, given the many unknowns regarding the structure of the accretion disc, and the complexity introduced by its ellipticity, we choose to focus solely on the companion star, and include a simple model for the accretion disc.

We used the \textsc{Icarus} code \citep{2012ApJ...748..115B} to create a model of the companion star in UW CrB, where irradiation of the companion star is allowed to vary over the duration of the optical burst. There are a few caveats to this model which we outline in the following paragraphs, before discussing our results.

First, in order to generate the companion's surface, the binary parameters for the system must be specified. As input, \textsc{Icarus} requires the following: the mass ratio of the binary ($q=M_{\rm NS}/M_{\rm Comp}$), the orbital period of the binary ($P_{\rm O}$), the inclination of the binary, $i$, the projected radial velocity amplitude of the companion star ($K_{\rm Comp}$), the co-rotation factor of the companion ($\Omega$), the Roche-lobe filling factor of the companion ($f$), the exponent of the gravity-darkening law ($\tau_{\rm grav}$), the base temperature of the companion star ($T_{base}$), the irradiating temperature affecting the companion star ($T_{irr}$), the distance modulus of the binary ($DM$), and the number of magnitudes of extinction in the direction of the binary in the Johnson V band ($A_v$).

Several of these parameters are not constrained for UW CrB and for which we make informed assumptions about in order to reduce the parameter space. First, the neutron star and companion star masses are unknown. As such, we assume a canonical neutron star mass of 1.55 M$_{\odot}$ (in line with the neutron star belonging to the recycled class of neutron stars which are found in LMXBs; \citealt{2016ARA&A..54..401O}), and allow the companion mass to vary between 0.1 and 0.3 M$_{\odot}$, in line with the expected companion mass for an accreting binary with an orbital period of 0.08 days. This leads to a mass ratio of between 5.17 and 15.5 respectively. We assume tidal locking ($\Omega=1$), and we fix the Roche lobe (RL) filling factor to 1, as we know that RL overflow is feeding the accretion disc. Finally, we fix the orbital period of the binary to 0.077067223 days, in line with the value obtained in this work.

For the accretion disc, we assume it is represented by the summation of a series of annuli, each radiating with a different black body temperature, and that the resultant spectrum is the sum of the contributions from each annulus. Each annulus produces black body radiation with a temperature given by $T(r)=T_{\rm in}(r/R_{\rm in})^{-0.5}$, where $T_{\rm in}$ is the temperature at the most inner part of the disc, $R_{\rm in}$. This profile assumes that the accretion disc is irradiated by the central source, as is expected in LMXBs \citep{1990A&A...235..162V,1996ApJ...464L.139V,1997ApJ...488...89K}. This parameterisation gives us three parameters to fit: the inner temperature of the accretion disc ($T_{\rm in}$), the ratio of the inner radius of the accretion disc to the outer radius, and a scaling factor ($S_{\rm g}$), which we define as the contribution of the disc to the observed $g'$ flux. 

For the inner radius of the disc, we assume that the disc extends to the surface of the neutron star, which we assume has a radius of 20 km. For the outer radius of the disc, we assume the disc extends to the tidal truncation radius the primary star's Roche lobe, which occurs at a radius of 0.9 $R_{\rm L}$ \citep{1977MNRAS.181..441P}.

Assuming this setup, there are two stages to the fitting. The first stage fits the pre-burst light curve in order to constrain the parameters $T_{\rm base}$, $T_{\rm irr}$, $i$, $DM$, $T_{\rm in}$, $S_{\rm g}$ and $A_v$. 

In the second stage, all of the these parameters were frozen other than $T_{\rm irr}$, and the burst light curve was fit. As such, for each data point of the burst, we calculate the contribution from the companion star assuming the binary parameters which are fit to the pre-burst light curve, and then optimise the irradiating temperature to minimise the difference between the observed flux and the expected contributions of the disc and the star.

\begin{figure*}
	\centering
	\includegraphics[width=1.0\textwidth]{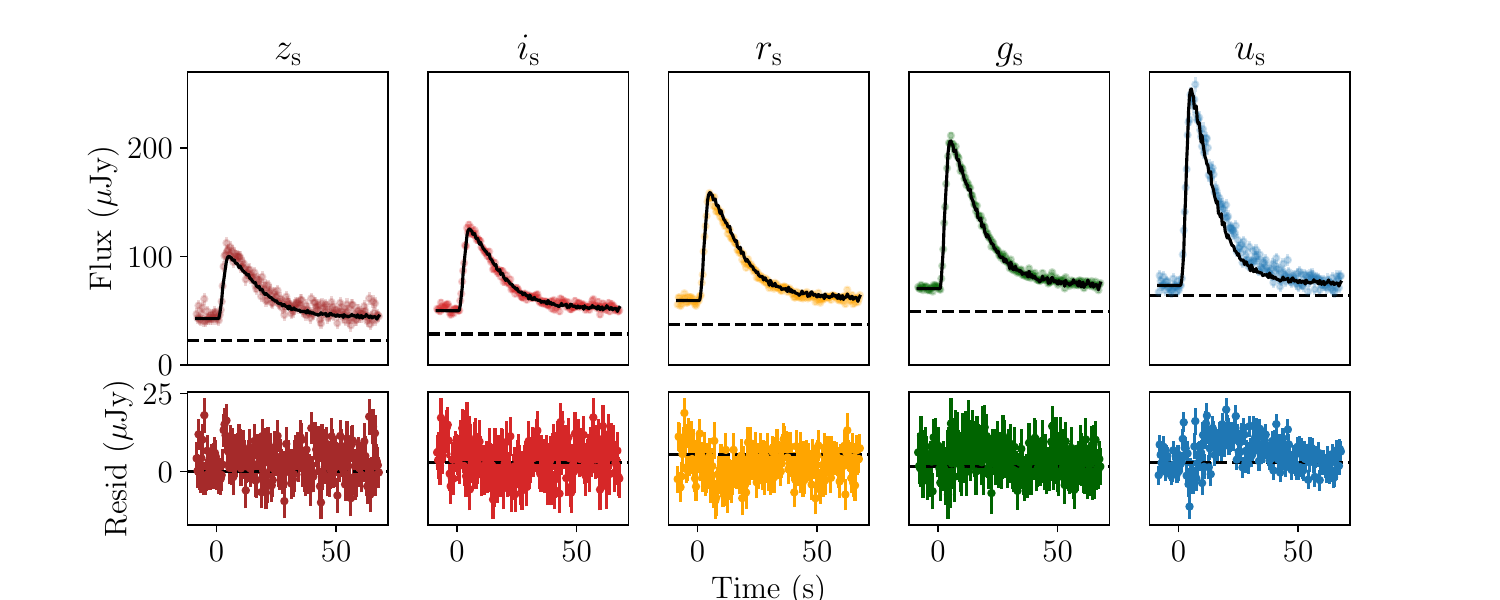}
	\caption{Best fitting \textsc{Icarus} model to the optical burst from UW CrB detected using HiPERCAM. Each panel shows the data from one of the filters (left to right: $z_{\rm s}$, $i_{\rm s}$, $r_{\rm s}$, $g_{\rm s}$, $u_{\rm s}$), along with the model as a solid black line. The dashed line shows the constant contribution to each band from the accretion disc. The x-axis is given in seconds relative to the start of the burst.}
	\label{fig:Icarus_model}
\end{figure*}

The parameter space was explored using \textsc{emcee} \citep{2013PASP..125..306F}, with 32 walkers allowed to evolve over 50,000 steps, with 1,000 steps were discarded as a burn-in. The corner plot of the resulting parameter space is included in Appendix~\ref{sec:corner_plots}. The best fitting parameters are given in Table~\ref{tab:Icarus_pars}, and the best fitting model shown in Figure~\ref{fig:Icarus_model}. There are a few caveats to our results. First, the temperature of the companion star and the irradiation it is subjected to outside of the burst are likely not physical constraints, as the values presented here assume the orbital variability in the light curve are all arising due to changing viewing aspects of the companion star. This is certainly not true, as the elliptical disc is likely contributing some fraction of the variability, especially if eclipses are occurring. However, while likely unphysical, the model is still useful for understanding the optical bursts. That is, the models show that increasing the temperature of the companion star dramatically can adequately model the burst (at least in most bands). The most obvious exception is the $u_{\rm s}$ band, where a significant deviation from the data can be seen during the decay of the burst. While we do not have a definite answer as to why the model diverges so poorly in this band relative to the others, it is worth noting that the Balmer jump occurs in this band. Given that Icarus uses photometry grids generated from LTE stellar atmosphere files, and the irradiated surface is likely not in LTE, the actual behaviour of the Balmer jump may be significantly different to the models we are using. 

It is also worth noting that accretion discs may have significantly different temperature and density gradients than stars. This is important as if the burst is primarily being reprocessed in the disc, then the line formation and evolution would be very different than in a star, which would also affect the Balmer jump. Hence, the poor fitting $u_{\rm s}$ band may be hinting at the reprocessing site lying in the disc, but further work and a more complex model is needed to test this idea.

\begin{table}
	\centering
	\caption{Best fitting binary parameters from the Icarus modelling of the HiPERCAM light curves}
	\begin{tabular}{l l}
		\hline
        Parameter               & Value\\
		\hline\hline
		$T_{\rm base}$ (K)	    & 6700$^{+700}_{-700}$\\
		$T_{\rm irr}$ (K)	    & 9600$^{+700}_{-500}$\\
		$T_{\rm in}$ (K)	    & $(8\pm1)\times10^{5}$\\
		$i$ ($^{o}$)	        & 68$^{+7}_{-13}$\\
		DM                      & 13.2$^{0.3}_{-0.2}$\\
		$S_{\rm g}$             & 0.70$\pm0.02$\\
		$A_{\rm v}$             & 0.62$\pm0.03$\\
        $M_{\rm Comp}$ (M$_{\odot}$) & 0.20$\pm0.07$\\
		\hline
	\end{tabular}
	\label{tab:Icarus_pars}
\end{table}

\begin{figure}
	\centering
	\includegraphics[width=1.0\columnwidth]{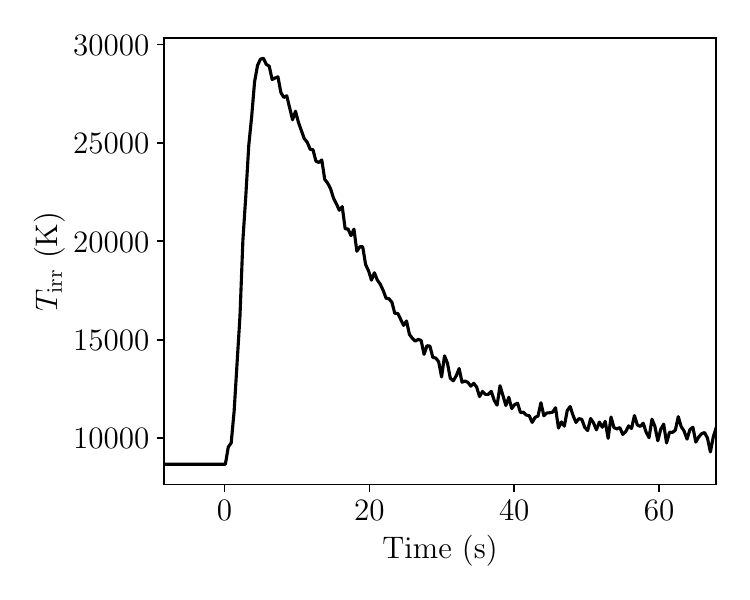}
	\caption{Evolution of the irradiating temperature parameter over the course of the burst for the best fitting model, showing that the peak temperature of the optical burst region reaches 28,000 K. The x-axis is given in seconds relative to the start of the burst.}
	\label{fig:Temp_Evo}
\end{figure}

Figure~\ref{fig:Temp_Evo} shows the evolution of the irradiating temperature over the course of the burst. Pre-burst, this value is at 9,300 K, suggesting the companion star is undergoing significant heating. Then, as the burst begins, it quickly rises to 28,000 K, followed by a slow decline. While this model can account for the overall shape of the $g_{\rm s}$, $r_{\rm s}$, $i_{\rm s}$, and $z_{\rm s}$ (up to a systematic offset which can be ascribed to a small photometric band calibration offset), the same cannot be said of the $u_{\rm s}$. This suggests that the burst spectrum may deviate from a black body at short wavelengths. This could be due to additional components which we are not accounting for, such as irradiation of the disc.

To improve on this model, a simultaneous optical and X-ray burst should be observed, with a time resolution similar to the HiPERCAM data presented here. Such a data set would allow for a lag to be measured, which would directly pinpoint the location of the reprocessing and allow for refinement of this model.

\subsubsection{Orbital phases of the bursts}\label{sec:PhaseOfBursts}
\begin{figure}
	\centering
	\includegraphics[width=\columnwidth]{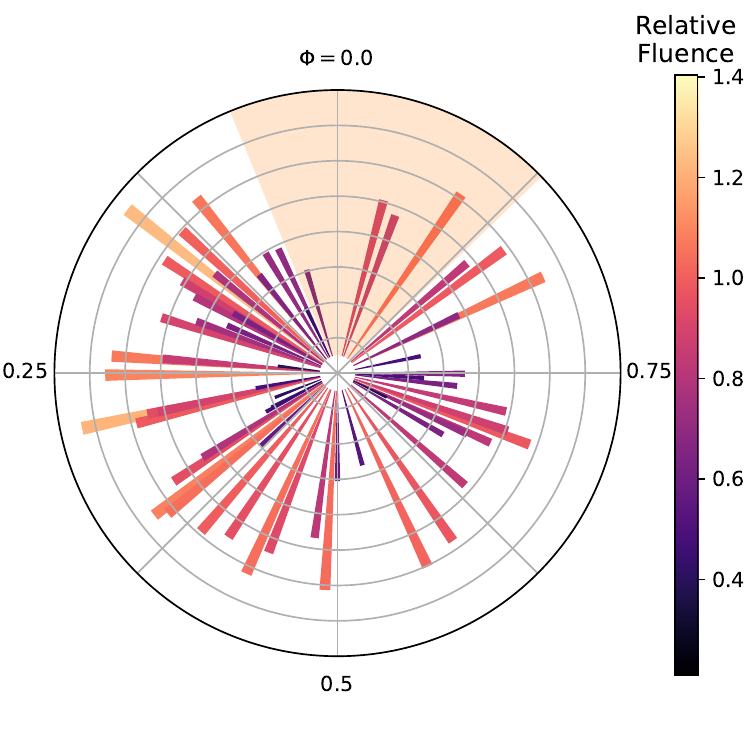}
	\caption[Phased bursts]{The fluences and orbital phases of every optical burst seen in UW CrB. This combines the 17 optical bursts presented in this work, the 4 UV bursts seen with the HST, the 33 optical bursts listed in \cite{2012AJ....144..108M}, and the 7 bursts shown in \cite{2015AcPPP...2...50M}. The orange shaded region shows the gap identified by \cite{2012AJ....144..108M}. There are now 4 detected bursts in this region.}
	\label{fig:phased_bursts}
\end{figure}

The relative fluence of all 17 optical bursts and the phases at which they occurred presented this paper were added to the 33 optical bursts listed in \cite{2012AJ....144..108M}, the 7 first presented in \citet{2015AcPPP...2...50M}, and the 5 bursts detected with the HST. Figure~\ref{fig:phased_bursts} shows a polar plot of the bursts, with the angle of the burst representing the orbital phase at which the burst occurred, and the radial extent of the burst representing the relative fluence of the burst. For this plot, we have normalised fluences of each burst by the median fluence observed with a given instrument+filter combination. This is to remove wavelength dependencies on the fluence, but we acknowledge it may be biasing the distribution if not enough bursts have been observed such that the observed median is not representative of the underlying, true median of fluences.

The gap in orbital phase during which no bursts had been observed by \cite{2012AJ....144..108M} now has 4 bursts present in it. \cite{2012AJ....144..108M} initially proposed that the apparent gap centred around orbital phase 0.967 was due to an eclipse of the reprocessing region, masking the reprocessing site from view. The detection of these two bursts rules out this theory. The four bursts had a relative fluence which was comparable to the average relative fluence of all the bursts, suggesting that the reprocessing site is not being eclipsed.

\section{The origin of the emission and absorption features.}
\begin{figure}
	\centering
	\includegraphics[width=\columnwidth]{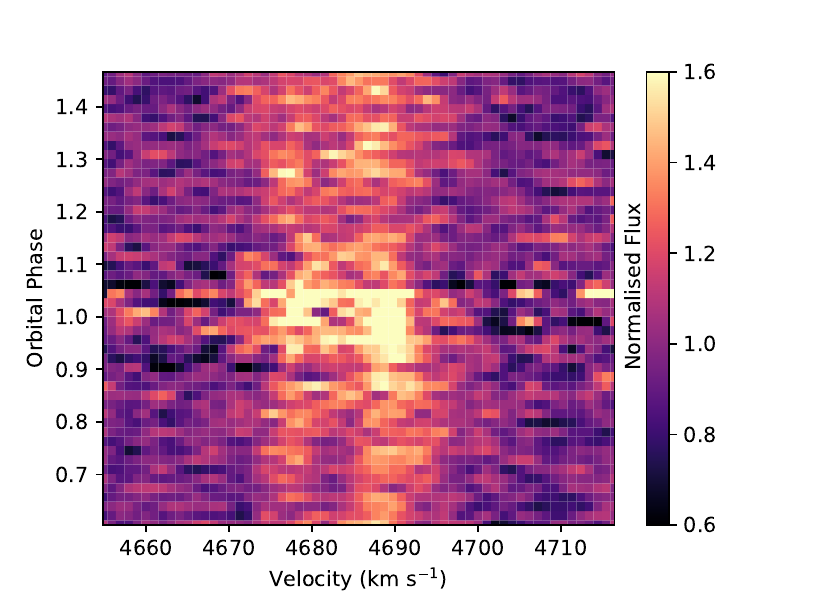}
	\caption{Trailed HeII 4686 \AA\ line after normalising to the continuum. These data show that the double peaked line persists throughout the eclipse, meaning it is not blocked by the secondary star.}
	\label{fig:trail_heII_norm}
\end{figure}

The optical minimum in UW CrB has been interpreted as the partial eclipse of the accretion disc in the system by the secondary star. This constrains the inclination of UW CrB to be close to 80\degree \citep{2008ApJ...685..428M}. The trailed optical spectra here support such an inclination for the system. HeII has a relatively high ionisation energy, meaning it must be coming from a location within the system which is substantially hotter than the location which produces the lower ionisation emission features of HeI and the Balmer features. Given the double peaked nature of the trailed spectrum, this region is likely to be within the inner part of the accretion disc. The lack of an eclipse in this line (only visible when the spectra are normalised to the continuum, see Figure~\ref{fig:trail_heII_norm}), could suggest that this region of the disc is not blocked at any time by the secondary star. 

However, this is not the only configuration which would produce a double peaked HeII line without an eclipse. Within other LMXBs, double peaked emission features have been detected even when the disc continuum has been eclipsed. Such behaviour has been proposed to be due to the origin site of the emission features lying above, but co-rotating with, the accretion disc \citep{2012A&A...539A.111S}. This is perhaps the case in UW CrB - in our spectra, the brightest emission line (H$\alpha$) is never fully eclipsed, and is only slightly reduced in flux during the eclipse. This behaviour in other systems suggests that inferring an inclination of 80\degree\: based solely on the behaviour of the emission features in the spectra would be unwise.

The behaviour of the HeI and Balmer lines suggest that these features are coming from a region which is in the outer part of the accretion disc. The orbital-phase dependant blue absorption features are consistent with an outflow from the system. Initial evidence for such an outflow has been recently proposed by \cite{2023MNRAS.526L.149F}, who found a transient, blue shifted absorption feature in far-UV spectra of UW CrB. However, the profiles of the features presented here are slightly different to those presented by \cite{2023MNRAS.526L.149F}. In the UV data presented in that paper, the emission features extend up to $-1000$ km s$^{-1}$ with a noticeable absorption dip embedded in the line profile at $-500$ km s$^{-1}$, while there is an additional absorption feature at an even higher velocity of $-1500$ km s$^{-1}$. In the data presented here, we only see an absorption feature at a velocity of $-500$ km s$^{-1}$, with no blue shifted emission and no higher velocity absorption feature. It may be that the cause of the $-500$ km s$^{-1}$ feature in the optical lines is responsible for the same feature in the UV data. However, this is likely not a disc wind, as these tend to cause features at much higher velocities in high-inclination systems ($>1000 {\rm \: km/s}$; see e.g. \citealt{2022A&A...664A.100P} Table 2, also  \citealt{2023A&A...679A..85C}), alike to the high-velocity component identified at UV wavelengths by \cite{2023MNRAS.526L.149F}.

The appearance and disappearance of the absorption feature present in our data is consistent with the orbital phase during which the impact region between the ballistic stream from the companion star and the outer edge of the accretion disc should be visible. If the impact region at the edge of the disc is inflated relative to the disc, then the absorption features could be a result of a wind originating closer to the NS interacting with this region. This would explain how a symmetric wind is giving rise to an absorption feature seen only at a particular viewing angle. Further phased resolved optical spectroscopy can be used to test this hypothesis, and repeated detection of absorption features are at this phase only would strengthen this scenario.

\section{Conclusions}
Four optical bursts have now been observed during the orbital phase gap observed by \cite{2012AJ....144..108M}, suggesting the site of reprocessing is not eclipsed for as long as previously thought. A narrow gap still exists at the predicted phase of inferior conjunction of the companion star. The existence of this gap permits the assumption assumed in this paper to model the bursts that all of the reprocessing occurs on the face of the secoundary star. However, if the gap is completely filled in, or a strong optical burst is detected during the middle of an eclipse, this would suggest the secondary star is responsible for very little of the reprocessed radiation, which motivate remodelling the data here but with a model that generates the reprocessed burst from a disc structure.

It also remains to be seen whether the presence of the orbital phase gap depends on the phase of the 5.5 day precessing accretion disc. As noted above, it is possible that the reprocessing site is located within the disc, so it is also possible that during certain parts of the 5.5 day precession of the disc, this site does become eclipsed during the orbital phases observed by \cite{2012AJ....144..108M} (logically, this would be when the site lies in between the NS and the secondary), and that during the other phases of the precessing disc, this site is not eclipsed (logically, this would be when the site lies on the far side of the NS, furthest from the companion). This argument is fully dependent on whether the reprocessing site is in the disc, which is yet to be confirmed. One thing is for certain - further optical monitoring of UW CrB, preferably over the 5.5 day disc period, should be performed to see if the detection of the bursts is dependant on the phase of the disc.

The optical spectra presented here raise more questions than they answer. Previously, the P Cygni profiles seen in HeI and H$\beta$ were thought to arise due to a wind from the NS. However, our time resolved spectra show that the blue absorption wing of these lines is transient, which is not expected for a wind component. It may be that our viewing angle of the accretion structures changes through the orbit, which leads to different amounts of absorption in the lines (this may also be related to the 5.5 day disc precession period). Furthermore, the brightest emission line (H$\alpha$) is not fully eclipsed, and is only slightly reduced in flux during the eclipse. This suggests the site of H$\alpha$ emission is not fully eclipsed by the secondary, and could mean that some emission is generated above the plane of the disc. Finally, we have seen evidence of the Bowen blend emission in the averaged spectrum. Further optical spectra with a high time resolution should be obtained to fully characterise the transient nature of the absorption in different spectral lines, and to track the Bowen blend emission over a full orbital period in order to constrain the mass of the NS.

Finally, we have applied a simple model of increased irradiation of the companion star to explain the amplitude and colour of the optical burst. While the model is simplistic, and ignores contributions from the accretion disc, it fits the data well, and shows that the luminosity of the burst is consistent with the inner face of the companion star reaching a temperature in excess of 25,000 K.

\section*{Acknowledgements}

We would like to sincerely thank the anonymous referee for the time and effort they put into their review. Their careful reading of the manuscript and thoughtful, considerate feedback lead to improvements in this paper, specifically in the modelling of the bursts.

M.R.K. and R.P.B. acknowledge support from the European Research Council (ERC) under the European Union's Horizon 2020 research and innovation programme (grant agreement No. 715051; Spiders). M.R.K acknowledges support from the Royal Society in the form of a Newton International Fellowship (NIF No. NF171019), and the Irish Research Council in the form of a Government of Ireland Postdoctoral Fellowship (GOIPD/2021/670: Invisible Monsters). P.A.M acknowledges support from Picture Rocks Observatory. I.P. acknowledges support from a Royal Society University Research Fellowship (URF\textbackslash R1\textbackslash 231496). D.M.S and T.M-D. acknowledge support from the Spanish \textit{Agencia estatal de investigaci\'on} via PID2021-124879NB-I00. D.M.S. also acknowledges support via a Ramon y Cajal Fellowship RYC2023-044941.

This work made use of \textsc{Astropy} \citep{astropy:2013,astropy:2018}, \textsc{corner} \citep{corner}, \textsc{emcee} \citep{2013PASP..125..306F}, \textsc{Icarus} \citep{2012ApJ...748..115B}.




\bibliographystyle{mnras}
\bibliography{uwcrb.bib} 



\appendix
\pagebreak
\section{Corner Plots}\label{sec:corner_plots}

\begin{figure*}
    \centering
    \includegraphics[width=\textwidth]{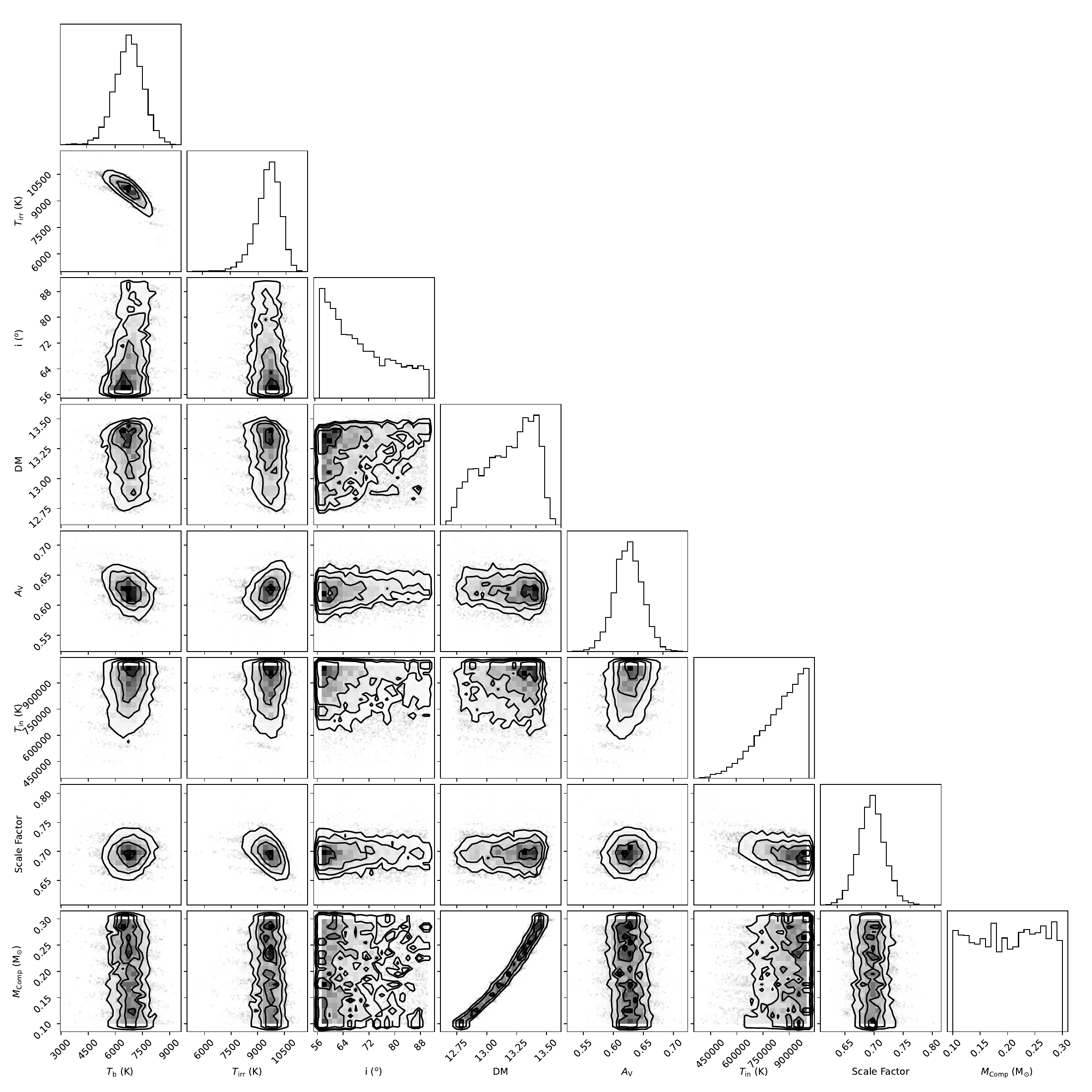}
    \caption{Corner plot from the MCMC analysis of the burst observed using HiPERCAM.}
\end{figure*}


\end{document}